\documentclass[aps,prl,twocolumn,amsmath,superscriptaddress,showpacs]{revtex4}

\usepackage{epsf}

\begin{document}

\title{Random Walks in Local Dynamics of Network Losses}

\author{I. V. Yurkevich}
\affiliation{School of Physics and Astronomy, University of
Birmingham, Edgbaston, Birmingham, B15 2TT, UK}

\author{I. V. Lerner}
\affiliation{School of Physics and Astronomy, University of
Birmingham, Edgbaston, Birmingham, B15 2TT, UK}

\author{A. S. Stepanenko}
\affiliation{School of Engineering, University of Birmingham,
Edgbaston, Birmingham, B15 2TT, UK}

\author{C. C. Constantinou}
\affiliation{School of Engineering, University of Birmingham,
Edgbaston, Birmingham, B15 2TT, UK}

\date{\today}

\begin{abstract}
We suggest a model for data losses in a single node of a
packet-switched network (like the Internet) which reduces to
one-dimensional discrete random walks with unusual boundary
conditions. The model shows critical behavior with an abrupt
transition from exponentially small to finite losses as  the data
arrival rate increases. The critical point is characterized by
strong fluctuations of the loss rate. Although we consider the
packet arrival being a Markovian process, the loss rate exhibits
non-Markovian power-law correlations in time at the critical
point.
\end{abstract}

\pacs{64.60.Ht, 
05.70.Jk, 
89.20.Hh, 
89.75.Hc 
}

 \maketitle


Many systems, both natural and man-made are organized as complex
networks of  interconnected entities: brain cells \cite{Arbib:01},
interacting molecules in living cells \cite{Jeong:00},
multi-species food webs \cite{Cohen:90}, social networks
\cite{Liljeros:01} and the Internet \cite{Pastor:01} are just a
few examples. In addition  to the classical Erd\"{o}s--R\'{e}nyi
model for random networks  \cite{ErdRen}, new overarching models
of scale-free \cite{Barabasi:99} or small-world \cite{Watts:98}
networks turn out to be describe real world examples. These and
other network models have received extensive attention by
physicists
 (see Refs.~\cite{Watts:99,Albert:02} for reviews).

One of particularly interesting problems for a wide range of
complex networks is their resiliency to breakdowns. The
possibility of random or intentional breakdowns of the entire
network has been considered in the context of scale-free networks
where nodes were randomly or selectively removed
\cite{Albert:00,Cohen:00,Braunstein:03}, or in the context of
small-world networks where  a random reduction in the sites
connectivity leads to a sharp increase in the optimal distance
across network which destroys its small-world nature
\cite{Braunstein:03,Dorogovtsev:00,Ashton:05}. In all these
models, the site or bond disorder acts as an input which makes
them very general and applicable to a wide variety of networks.

Network breakdowns    can  result not only from a physical loss of
connectivity but  from an operational failure of some network
nodes to forward data. In the more specific class of communication
networks, this could happen due to excessive loading of a single
node. This could trigger cascades of failures and thus isolate
large parts of the network \cite{Moreno:03}. In describing the
operational failure in a particular network node, one needs to
account for distinct features of the dynamically `random' data
traffic which can be a reason for such a breakdown.

In this Letter we model data losses in a \textit{single node} of a
packet-switched network like the Internet. We demonstrate that
such losses may have critical behavior with an abrupt transition
from an exponentially small to finite loss rate as  the data
arrival rate increases. At the critical point the loss rate
exhibits strong fluctuations which only become Gaussian in the
(unrealistically) long time limit.  Although we model data
arrivals as a Markovian process, the loss rate at intermediate
times shows long-range power-law correlations in time. When
excessive data losses start, it is more probable that they persist
for a while, thus impacting on network operation.

There are two distinct features which must be preserved in
modelling data losses in a packet-switched network: a discrete
character of data propagation and the possibility of data overflow
in a single node.  In such a network, data is divided into packets
which are routed from source to destination via a set of
interconnected nodes (routers). At each node packets are queued in
a memory buffer before being {\it served}, i.e.\ forwarded to the
next node. (There are separate buffers for incoming and outgoing
packets but we neglect this for the sake of simplicity). Due to
the finite capacity of memory buffers and the stochastic nature of
data traffic, any buffer can become overflown which results in
discarded packets.

In the model we suggest here data losses in a single memory buffer
start when the average rate of random packets arrival exceeds the
service rate. In such a model the transition from free flow to
lossy behavior is, \textit{on average}, very steep: when the
arrival rate exceeds a certain threshold, the buffer becomes full
and a finite fraction of arriving packets is dropped. Such a sharp
onset of network congestion is familiar to everyone using the
Internet and was numerically confirmed in different models
\cite{Ohira:98}. Here we stress two characteristic consequences of
the model considered which would be preserved in any realistic
model allowing for the discrete data propagation and finite
capacity of the  nodes: (i) congestion can originate from a single
node and (ii) loss rate statistics turns out to be highly
nontrivial. The latter makes present considerations qualitatively
different from, e.g., bulk queue models which have been
extensively studied before \cite{Cohen:69,Schwartz:87}  but
considered loss rate only on average. Although fluctuations in
network dynamics were studied in \cite{Menezes}, this was done in
a continuous limit for the data traffic. In our model there is a
more close analogy with mesoscopic physics where, e.g., the
electron density of states in disordered conductors is, on
average, a constant, but its fluctuations are rather nontrivial,
either globally or locally \cite{AS}.

We consider  the  model of randomly arriving packets   which  form
a queue in the buffer and are served at regular, discrete time
intervals. The length of the queue after $n$ service intervals,
$\ell_n$ serves as a dynamical variable. The random arrival is
modelled as a telegraph noise described by the discrete-time
Langevin equation,
\begin{align}\label{L}
    \ell_{n+1}=\ell_{n}+\xi_{n}\,,
\end{align}
where  the simplest model for noise $\xi_{n}$ is built by assuming
that all incoming packets are of the same length  and arrive one
by one at each service interval with probability $p$. We further
assume that one half of a packet is served at each interval on a
first-come first-served basis. Let $L$ be the buffer capacity,
i.e.\ the maximal  number of service units (half-packets) in the
queue. This gives
\begin{align}\label{1}
    \xi_n= \left\{%
\begin{array}{lc}
    1, &{0\le\ell_n\le L-1} \\
    0, &{\ell_n=L}\,, \\
\end{array}%
\right.
\end{align}
 with  probability $p$, and
\begin{align}\label{2}
    \xi_n= \left\{%
\begin{array}{lc}
       \phantom-0, &{\ell_n=0}\,, \\
 -1, &{1\le\ell_n\le L} \\
\end{array}%
\right.
\end{align}
with  probability $1-p$. The meaning of the above conditions is that
either the length of the queue increases by one service unit when one
packet arrives and one service unit is served, or decreases  by one
 when no new packet arrives. The approximate boundary
conditions above correspond to discarding a newly arrived packet
when buffer is full ($\ell_n=L$) and an idle interval when no
packet arrives at an empty buffer ($\ell_n=0$). We  show at the
end of the Letter that more accurate boundary conditions do not
change the asymptotic form of our results.

\begin{figure}[t]

\begin{center}
\leavevmode \epsfxsize=0.4\textwidth \epsffile{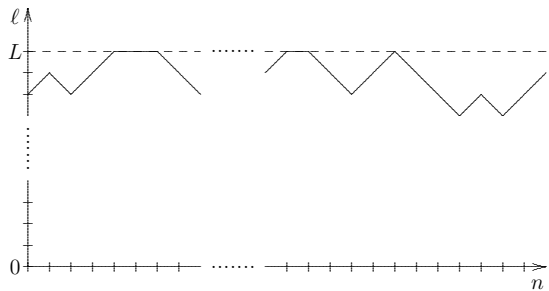}
\end{center}

\caption{The model of data losses: incoming packets randomly arrive in
discrete time intervals and join the queue of length $\ell$ limited by the
memory buffer capacity $L$. Packets in front of the queue are served at
the same time intervals. If the queue reaches the ``boundary'' (the buffer
is full), newly arriving packets are discarded.}

\label{Figure 1}

\end{figure}

The main quantity which characterizes congestion is the packet loss rate
which is defined via the number of packets discarded during a time
interval $N$ by
\begin{align} \label{loss}
\mathcal{L}_N(n_0)= \sum_{n=n_0+1}^{n_0+N} \delta_{\ell_n,
L}\delta_{\ell_{n+1}, L}\,.
\end{align}
The meaning of this definition is that the arriving packet is discarded
when by the moment of arrival the queue was at the maximal capacity $L$ as
illustrated in Fig.~\ref{Figure 1}. Thus the continuous limit cannot be
exploited for this problem which makes the loss statistics profoundly
different from, e.g., the thoroughly studied statistics of first-passage
time \cite{Hughes-RandWalk:95}. We will find the average and the variance
of the loss rate defined above. Although the arrival of packets defined by
the Langevin dynamics of Eqs.~(\ref{L})--(\ref{2}) is a Markovian process,
we will show that the loss rate dynamics turns out to be non-Markovian in
the critical regime. The reason for this is that the loss rate
(\ref{loss}) is defined entirely by the process occurring at the boundary
of the random walk (RW).

We will express the quantities of interest via the conditional probability
of the queue being of length $\ell$ at time $n$ provided  that it was of
length $\ell'$ at time $ n_{0}$,
\begin{align*}
{\mathcal{G}}_{n-n_0}({\ell},{\ell}')=\left<{\delta_{{\ell}_n
,{\ell}}\, \delta_{\ell_{n_0},{\ell}'}}\right>/\left<{
\delta_{\ell_{n_0},{\ell}'}}\right>\,,
\end{align*}
where $\left<{\ldots}\right>$ stand for the averaging over the
telegraph noise of Eqs.~(\ref{L}) -- (\ref{2}).  The stationary
distribution of the queue length is related to ${\mathcal{G}}$ by
\begin{align} \label{st}
    \mathcal{P}_{\text{st}}(\ell)=\lim_{n_0\to-\infty}
    {\mathcal{G}}_{n-n_0}(\ell,\ell')=
\left<{\delta_{{\ell}_n ,{\ell}}} \right>\,.
\end{align}
On averaging the loss rate, Eq.~(\ref{loss}), we thus obtain
\begin{align}
     \label{l-av}
    \left<{{\mathcal{L}}_N}\right>=
 \mathcal{P}_{\text{st}}(L)N{\mathcal{G}_1(L,L)}\,,
\end{align}
while its variance is given by
\begin{align}
     \label{l-var}
    &\left<{{\mathcal{L_N}}^2}\right>= \sum_{n,m=n_0+1}^{n_0+N}\left<{
    \delta_{\ell _n,L} \delta_{\ell _{n+1},L}
   \delta_{\ell _m,L} \delta_{\ell _{m+1},L}
    }\right>\notag\\
    &=\left<{{\mathcal{L}}_N}\right>+2{\mathcal{P}_{\text{st}}}(L){\mathcal{G}}_1^2
    (L,L)\sum_{n<m}^{}{\mathcal{G}}_{m-n-1}(L,L)\,.
\end{align}

To calculate ${\mathcal{G}}$ (which is non-trivial due to the boundary
conditions of Eqs.~(\ref{1}) and  (\ref{2})), we note that it is the
Green's function of the Focker-Planck equation (FPE) corresponding to the
Langevin equation (\ref{L}). The FPE can be written in terms of the
probability ${\mathcal{P}}_n(\ell )$ for the queue being of length $\ell $
at time $n$ as
\begin{align} \label{P}
    {\mathcal
P}_{n+1}(\ell)&=\sum_{\ell'}w_{\ell,\ell'}{\mathcal P}_n(\ell')\,,
&&0\le\ell,\ell' \le L\,.
\end{align}
The transition matrix $\hat w$ with elements $w_{\ell,\ell'}$
corresponding to Eqs.~(\ref{L})--(\ref{2}) is given by
\begin{align}\label{w}
w_{\ell,\ell'}&=p\,\delta_{\ell-1,\ell'}+(1-p)\,\delta_{\ell+1,\ell'}\,,
&&0<\ell  <L\,,
\end{align}
with the boundary conditions
\begin{align}\label{w0}
   w_{\ell,\ell'}= \left\{%
\begin{array}{rl}
    (1-{p})\bigl(\delta_{0,\,\ell '}
  +\delta_{1,\,\ell '}\bigr), & \ell=0 \\[4pt]
    {p}\,\bigl(\delta_{L-1,\,\ell '} +\delta_{L,\,\ell
'}\bigr), & \ell=L \\[4pt]
    (1-{p}) \delta_{ \ell \,,0} +p\,\delta_{ \ell\,,1  }, & \ell'=0\\[4pt]
    (1-{p}) \delta_{\ell,\, L-1 } +{p}\,\delta_{ \ell\,,L}, &
    \ell'=L \\
\end{array}%
\right.    \,.
\end{align}
Eqs.~(\ref{P})--(\ref{w}) describe a  usual biased discrete-time RW  on a
one-dimensional lattice \cite{Hughes-RandWalk:95}. However, both the
quantity to calculate, Eq.~(\ref{loss}), and the boundary conditions,
Eq.~(\ref{w0}), make the problem under consideration  profoundly different
from those in \cite{Hughes-RandWalk:95}.

Eqs.~(\ref{P}) -- (\ref{w0})  are clearly non-Hermitian. This
leads to different right, $\psi^+$,  and left, $\psi^-$,
eigenfunctions of the
  matrix $\hat w$ (normalized by
  $\sum_{\ell=0}^{L}\psi^+(\ell)\psi^-(\ell)=1$):
\begin{align} \label{RL}
    {\hat w}\psi^+_k&=\lambda_k\,\psi^+_k\,,&{\hat
    w}^T\psi^-_k&=\lambda_k\,\psi^-_k\,,
\end{align}
where $\lambda_k$ are the eigenvalues, labeled with a discrete `momentum'
$k$. Although there exists a similarity transformation which turns the
problem into  Hermitian  (which means that all $\lambda_k$ are real),  it
is convenient to keep the above representation unchanged.

The Green's function  of the FPE (\ref{P}) can immediately be expressed as
${\hat{\mathcal{G}}}_n=\hat w^n$ which gives
   \begin{equation*}
    {\mathcal{G}}_{n}(\ell ,\ell ')=\sum_k\lambda_k^n\psi^+_k(\ell )\psi^-_k(\ell ')\,.
\end{equation*}
Diagonalizing   the tri-diagonal matrix $\hat w$ defined by Eqs.~(\ref{w})
and (\ref{w0}), one finds the eigenvalues of Eq.~(\ref{RL})
\begin{align}\label{eigv}
    \lambda_k&= 2\sqrt{p(1-p)}\cos k\,,
\end{align}
where $k=\pi n/(L+1)$, $n=1,2,\ldots L$. The appropriate
eigenfunctions are given by
\begin{equation} \label{R-eig}
    \begin{aligned}
    \psi^\pm_k(\ell )&=c_k q^{ \pm\ell/2 }\left[\sin k(\ell +1)-
        q^{ 1/2 }\sin k\ell \right]\,,\\
  c_k^2&=\frac{2p}{(L+1)(1-\lambda_k)}\,,\quad q\equiv \frac{p}{1-p}\,.
\end{aligned}
\end{equation}
The eigenfunctions corresponding to $k=0$ are given by
\begin{align}\label{0}
\psi^+_0(\ell )&=c_0
q^{\ell}\,,&\psi_0^-(\ell)&=c_0\,,&&c_0^2=\frac{1-q}{1-q^{L+1}}\,,
\end{align}
and the appropriate eigenvalue, $\lambda_0=1$ is separated by a gap from
the continuous (as $L\to\infty$) spectrum of Eq.~(\ref{eigv}), unless
$p=1/2$. The RW is biased towards $\ell\to L$ (full buffer and congested
traffic) for $p>1/2$, or towards $\ell\to0$ (empty buffer) for $p<1/2$. At
$p=1/2$  when the  RW is unbiased and the eigenvalue spectrum is gapless,
the fluctuations are strongest.  In all cases, since $\lambda_0=1$ while
$\lambda_k<1$ for $k\ne0$, it is the isolated solution (\ref{0}) which
governs the stationary  distribution (\ref{st}):
\begin{align}\label{st2}
    \mathcal{P}_{\text{st}}(\ell)=
    \lim_{n_0\to-\infty}{\mathcal{G}}_{n-n_0}(\ell,\ell')=c_0^2q^\ell\,.
\end{align}

Noticing that ${\mathcal{G}}_1(\ell ,\ell ')=w_{\ell ,\ell '}$ so
that ${\mathcal{G}}_1(L,L)=p$, one finds the average loss rate
from Eqs.~(\ref{l-av}) and (\ref{st2}) as
\begin{align*}
     \frac{1}{N}\left<{{\mathcal{L}}_N}\right>&=
    p\frac{q^{L+1}-q^L}{q^{L+1}-1} \;  \begin{array}{c}
        \\[-8pt]
      \longrightarrow \\[-8pt]
      _{L\gg1} \\
    \end{array}\;
    \left\{%
\begin{array}{ll}
    2p-1, & {p>\frac12;} \\[6pt]
    \frac{1}{L+1}, & {p=\frac12;} \\[6pt]
    \frac{1-2p}{1-p}\,q^L, & {p<\frac12,} \\
\end{array}%
\right.
\end{align*}
so that the loss rate is a constant of order 1 for $p>1/2$, a small
fraction of the buffer capacity for $p=1/2$ and an exponentially vanishing
fraction for $p<1/2$. This straightforward result could be obtained
directly from the Langevin description. The matching between these three
asymptotic regimes takes place in a narrow region (of width $\sim1/L$)
around $p=\frac{1}{2}$.

\begin{figure}[t]

\begin{center}
\leavevmode \epsfxsize=0.45\textwidth \epsffile{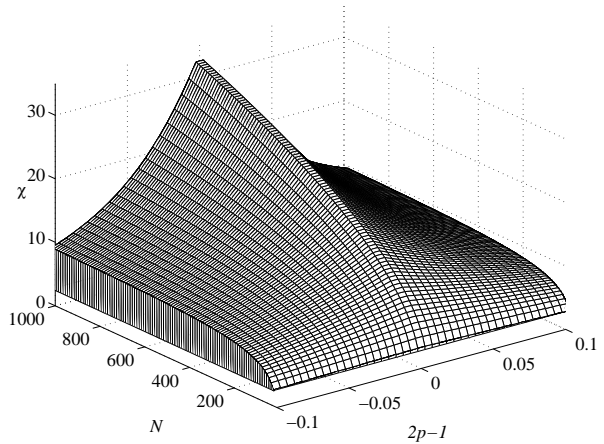}
\end{center}

\caption{compressibility $\chi$ (for $L=1000$) shows a fast increase of
fluctuations in time  at the critical point, $p=\frac{1}{2}$.  }

\label{Figure 2}

\end{figure}

The result for the variance, Eq.~(\ref{l-var}), is convenient to
express in terms of the `compressibility' defined by
\begin{align} \label{var-chi}
     \langle \delta{\cal L}^2_N\rangle&\equiv\chi_N\,\langle
{\cal L}_N\rangle\,, &\delta\mathcal{L}_N(n) &= \mathcal{L}_N(n)-
\left\langle
     \mathcal{L}_N\right\rangle\,.
\end{align}
From Eqs.~(\ref{l-var}) and (\ref{R-eig}) we find
\begin{align}\notag
     \chi_N&= 1{ -p\,\mathcal{P}_{\text{st}}(L)} \\&+
\frac{4p\,(1-p)}{L+1}\sum_{k>0}\frac{\sin^2k}{(1-\lambda_k)^2}
\label{L-var}
   \left[1-\frac{1}{N}\frac{1-\lambda^N_k}{1-\lambda_k}\right].
\end{align}
The behavior of $\chi$ is illustrated in Fig.~\ref{Figure 2} which shows
its fast increase at the critical point, $p=1/2$. Using  Eq.~(\ref{eigv})
for $\lambda_k$, it is easy to simplify Eq.~(\ref{L-var}). We find  that a
steady-state regime  (when one  neglects the $N$-dependent term in the
square brackets above) is reached for $N\gg N_0$ where
\begin{align*} 
     N_0\equiv \left[(2{p}-1)^2+(\pi/L)^2\right]^{-1}\,.
\end{align*}
In this regime, the compressibility saturates at
\begin{align*}
    \chi_\infty=\left\{%
\begin{array}{ll}
    \frac{1-|2p-1|}{|2p-1|}, &  |2p-1|L\gg1 \\[8pt]
    \frac23L,&|2p-1|L\ll1 \\
\end{array}%
\right.
\end{align*}
Thus, the compressibility diverges at the transition point $p=1/2 $ in the
thermodynamic limit, $L\to\infty$ and $N/L^2\to \infty$.  The variance
(\ref{var-chi}) remains finite at the transition point and in the
thermodynamic limit it obeys the central limit theorem.

However, at the critical point, $p=1/2$, the steady-state regime is
reachable only at unrealistically long times $N\!\gg \!N_0\!\propto\!
L^2$. In the intermediate regime, $1\ll N\ll N_0$, the compressibility
rapidly increases with time:
\begin{align} \label{chiN}
\chi_{N}&=c\,N^{1/2}, & c&=\frac{2\sqrt
2}{\pi}\int\limits_{0}^{\infty}\frac{dx}{x^2}\left(1-\frac{1-e^{-x^2}}{x^2}\right),
\end{align}
so that the variance exceeds the average value of the loss-rate
and its distribution is no longer normal. More importantly, in
this regime the fluctuations of the loss rate are no longer
Markovian as they exhibit long-time correlations. To show this,
we consider the temporal correlation function of the loss rate
defined by
\begin{equation*}
     R_2(N,M)\equiv
     \frac{\left\langle \delta\mathcal{L}_N(0)\delta\mathcal{L}_N(M)\right\rangle}%
     {\left\langle \delta\mathcal{L}_N^2\right\rangle}\, ,\ \ M>N\,.
\end{equation*}

We obtain an exact expression for $R_2(N,M)$ similarly to that for
$\chi_N$, Eq.~(\ref{L-var}), omitted for brevity. In the most relevant
regime, $N_0\gg N\gg1$ and $M>N$, it reduces to
\begin{align}\notag
     R_2(N,M) = &\frac{p N}{\chi_N}
     \Biggl[ {\rm e}^{-M(2p-1)^2/2}\sqrt{\frac{2}{\pi M}}\\
     &
     - |2p-1|\mathop{\rm erfc}\left(|2p-1|\sqrt{M/2}\right)\Biggr].
\end{align}
At the critical point this reduces using Eq.~(\ref{chiN}) to
\begin{equation}\label{R2}
     \left.R_2(N,M)\right|_{p=1/2} = c^{-1}\sqrt{\frac{N}{2\pi M}}\,.
\end{equation}
This long-time correlation (in spite of the packet arrival being
Markovian) is another clear sign of criticality.

Let us note that the boundary conditions in Eq.~(\ref{w0})
correspond to simultaneous arrival and service of packets. In this
case overflown packets are only partially discarded. In more
realistic models the overflown packets should be discarded
completely. To reflect this, we can choose one of the standard
procedures: service first or packet arrival first. This is
straightforward to formulate: the  transition matrix remains the
same in the bulk, Eq.~(\ref{w}), while changes in $3\times 3$
blocks in the boundary corners. In  solving the eigenvalue problem
(\ref{RL}) the appropriate boundary layer states can be
eliminated. This reduces our problem to that described by
Eqs.~(\ref{w}) and (\ref{w0}) but with a smaller number of states
and different (and dependent on eigenvalues) corner elements  on
the main diagonal. This can be solved in a similar way as   the
model of Eqs.~(\ref{P})--(\ref{w0}) and the dependence on $N$ and
$M$ turns out to be the same in the asymptotical regime of
Eqs.~(\ref{chiN})--(\ref{R2}).

In conclusion, we have demonstrated that the stochastic nature of
discrete data traffic in  packet-switched networks (e.g., the
Internet) results in a critical behavior with an abrupt transition
from free to lossy operation at the level of a single node when
the arrival rate reaches a certain critical value.
  The critical point is characterized
by strong fluctuations and long-memory effects in the loss rate.
This leads to an operational failure of a single node which can
contribute to cascaded failures and thus congestion of large parts
of the network. We intend to use the results of the present model
as building blocks for describing such a congestion within the
framework that accounts also for the topological disorder
\cite{Stepanenko:05}.

\acknowledgements{This work was supported by the EPSRC grant
GR/T23725/01.}

 \end{document}